\documentclass[%
prd
 ,secnumarabic%
,amssymb, amsmath,nobibnotes,showpacs]{revtex4}
\usepackage{bm}%
\usepackage{graphics}
\expandafter\ifx\csname package@font\endcsname\relax\else
 \expandafter\expandafter
 \expandafter\usepackage
 \expandafter\expandafter
 \expandafter{\csname package@font\endcsname}%
\fi

\begin{document}

\title{Reply to ``Revisiting the solution of the second-class constraints of the Holst action''}%

\author{Francesco Cianfrani$^{1}$, Giovanni Montani$^{23}$}%
\email{Francesco.Cianfrani@uniroma2.it; giovanni.montani@enea.it}
\affiliation{$^1$ University of Rome Tor Vergata, Department of Industrial Engineering, Via del Politecnico 1, Rome 00133, Italy\\
$^2$ ENEA, Fusion and Nuclear Safety Department, C. R. Frascati, Via E. Fermi 45, 00044 Frascati (Roma), Italy\\
$^3$ Physics Department, ``Sapienza'' University of Rome, P.le Aldo Moro 5, 00185 Roma (Italy)}

\begin{abstract}
Reply to arXiv:1903.09201.
\end{abstract}

\pacs{04.60.Pp, 11.30.Cp}

\maketitle

The authors of \cite{Montesinos:2019ypp} completely misunderstood the results of our manuscript \cite{Cianfrani:2008zv}. They claim that we ``provided a certain parametrization of the solution of the second-class constraints resulting in a noncanonical symplectic structure (for which they also asserted to have found canonical variables). Nevertheless, such a symplectic structure is incorrect because it involves 21 variables and thus does not give the right count of the local degrees of freedom (d.o.f.) of general relativity.''

Indeed, we provided a parametrization of the phase space in terms of 24 variables, so providing the right number of degrees of freedom of General Relativity. These variables are: 
\begin{itemize}
\item the connections $\tilde{A}^a_i$ and their conjugate momenta $\tilde{\pi}^i_a$, that reduce to Ashtekar-Barbero connections and inverse densitized triads in the time-gauge, 
\item the boost variables $\chi_a$ and their conjugate momenta $\tilde{\pi}^a$. 
\end{itemize}
 
This is clear by looking at Eq.(12) of the paper, in which the action is written in terms of these variables. 

Therefore, the motivation for the analysis given in \cite{Montesinos:2019ypp} is inconsistent.

\end{document}